**Title**
Directional picoantenna behavior of tunnel junctions in the presence of atomic-scale defects.

**Authors**
David Mateos[1,2,†], Oscar Jover[1,2,†], Miguel Varea[1,2,†], Koen Lauwaet[1], Daniel Granados[1], Rodolfo Miranda[1,2], Antonio I. Fernandez-Dominguez[3], Alberto Martin-Jimenez[1*], Roberto Otero[1,2].

**Affiliations**
1. IMDEA Nanociencia, Madrid, Spain.
2. Departamento de Física de la Materia Condensada and Condensed Matter Physics Center (IFIMAC), Universidad Autónoma de Madrid, Madrid, Spain.
3. Departamento de Física Teórica de la Materia Condensada and Condensed Matter Physics Center (IFIMAC), Universidad Autónoma de Madrid, Madrid, Spain.

[*] Corresponding author.
[†] Equally contributing authors.

**Abstract**
Plasmonic nanoantennas, metallodielectric structures with engineered size and shape, have attracted much attention lately as they make the control of the directionality and temporal characteristics of light emitted by fluorophores possible. Nanoantennas exploit light-matter interactions mediated by Localized Surface Plasmon Resonances and, so far, have been demonstrated using metallic nanoparticles or other metallic nanostructures. Plasmonic picocavities, i.e., plasmonic cavities with mode volumes below 1 cubic nanometer, could act as antennas to mediate light-matter interaction even more efficiently than their nanoscale counterparts due to their extreme field confinement, but the directionality on their emission is difficult to control. In this work, we show that the plasmonic picocavity formed between the tip of a Scanning Tunnelling Microscope and a metal surface with a monoatomic step shows directional emission profiles and, thus, can be considered as a realization of a picoantenna. Comparison with electromagnetic calculations demonstrates that the observed directionality arises from light emission tilting of the picocavity plasmons. Our results, thus, pave the way to exploiting picoantennas as an efficient way to control light-matter interaction at the nanoscale.

**Teaser**
We have fabricated a picoantenna by approaching a metal tip to a plasmonic surface close to a naturally occurring atomic-scale defect known as a step edge. We demonstrate that, by injecting current through the created tunnel junction, we can excite the plasmonic modes of the picocavity, and the radiation arising from their radiative relaxation can be observed in the far field. By studying step edges with different orientations, we demonstrate that the light emission is directional and, thus, that such picocavity acts as a picoantenna, opening new avenues to control light-matter interaction at atomic scales.

**MAIN TEXT**

**Introduction**
Manipulation of light at subwavelength scales can be accomplished by leveraging the extreme confinement of electromagnetic fields by metallic nanostructures, supporting Localized Surface Plasmon (LSP) resonances. The size and shape of the nanostructure ultimately determine the near-field interaction between LSPs and nanoscale light sources, such as atoms, molecules, or quantum dots, which in turn dictates the radiated power and spectral and angular distributions of the emitted light in the far field (*1–4*). In particular, dielectric gaps are a fundamental element in engineered metallic nanostructures, which can be designed to shape the emission of fluorophores they host, and, in this context, they are usually termed nanoantennas. So far, control over the directionality of light has been achieved at the scale of several nanometers. Different strategies involving single nanoparticles (*5–7*) and nanoparticle arrangements (*8–14*) have been proposed to manipulate the directionality of light at the subwavelength scale.

In the quest for stronger light-matter interactions, it has been shown that light can be confined into even smaller dimensions when atomic features are present on the nanocavity gap, giving rise to so-called "picocavity" plasmons (*15–19*). Picocavity plasmons confine light below 1 cubic nanometer, which pushes the optical spatial resolution down to the sub-nanometer scale (*20, 21*). Controllable picocavities are, however, hard to achieve due to the inherent lack of stability and reproducibility of the atomic-scale features in the gap, where very high electromagnetic fields can be present. Such control over atomic-scale features is, however, routinely achieved in Scanning Tunnelling Microscopy (STM). STM allows for the fabrication, manipulation, and imaging of structures consisting of just a few atoms, where atomic scale features of metallic surfaces, such as monoatomic step-edges, can be studied even at room temperature. Moreover, the gap between the metallic STM tip and the sample surface is known to act as a metallic nanocavity, hosting LSP modes that can be excited by inelastic tunneling electrons. Positioning the tip at different positions with respect to an atomic-scale feature of the surface, thus, leads to tunable picocavities, which could act as picoantennas that, hopefully, might allow for the control in the directionality of the emitted light.

In this work, we analyze the plasmonic response of a metallic STM picocavity comprised of a Au tip and a Ag(111) sample across a monoatomic height step. We observe that, upon removal of the electronic factors from the raw optical spectra (*22*), the radiated intensity at the step edge can be larger or smaller than that arising from flat terraces, depending on the orientation of the step edge to our light collection angle. This effect can be observed regardless of the particular tip used to create the cavity. The origin of these effects can be traced by comparison with electromagnetic calculations. According to our calculations, the angular distribution of the emitted light is isotropic when the tip is far from the step (>3 nm), but it becomes strongly anisotropic when the tip is near the step (+/- 1 nm), leading to larger (lower) intensities being radiated towards the half-plane that contains the upper (lower) terrace. Thus, depending on the orientation of the step edge, the directional nature of the emitted light results in a better/worse alignment to our collection setup (that is fixed), resulting in an increase/decrease of the detected light in the far field, in good correspondence with our experiments. Moreover, the spectral redistribution obtained in our experiments corresponds closely to that expected from theoretical calculations. These results clearly demonstrate that a metal STM tip in close proximity to an atomic-scale defect can be considered as a realization of a picoantenna, capable of performing its most characteristic function, that is, controlling the directionality of the emitted light. We can envision that this technique could be exploited to tune the direction of light emission of molecules, quantum dots, or other single quantum emitters, opening a new window to investigate the optical properties of atom-size objects.

**Results**

A schematic representation of the experimental setup is illustrated in Figure 1A. The light emitted by the radiative relaxation of the LSP modes of the tip-sample picocavity excited by inelastically tunneling electrons is collected by a fixed lens. The corresponding STM image shows the surface of a Ag(111) sample with several monoatomic height steps. A typical broadband raw plasmonic spectrum of the picocavity formed by a Au tip and an atomically clean Ag(111) sample is displayed by the red curve in Figure 1B, where its specific structure (number of subpeaks and intensity ratios) depends on the shape of the STM tip (*23, 24*). As the black and blue curves in Figure 1B show, the total emitted light is dramatically reduced when the tip is placed atop a monoatomic height step edge. Figure 1C displays a series of horizontally stacked spectra traversing two steps across the black dotted line in Figure 1A, where the color scale represents the raw light intensity. Notice that the scanned line consists of a first step in downward orientation and a second upward, with respect to our collection lens (our collection angle is oriented along the negative horizontal axis in Figure 4A). The color-coded image displays two abrupt light intensity drops that, by comparison with the height profile of the lower panel, can be attributed to the presence of the steps. A comparison of both graphs shows that the measured light intensity reduction is extremely local and occurs over a lateral extension of <1 nm near each step. These results are in good agreement with previous reports on the effect of step edges and other atomic-scale defects on Scanning Tunneling Microscopy-induced Light emission (STML) spectra (*25–28*). Several interpretations have been proposed for such a phenomenon, from variations in the electromagnetic coupling between tip and sample at step edges (*26*) to the effect of the modified electronic density of states at the step edges, which will affect the number of available final states for inelastic tunneling (*27, 28*) and perhaps affect the branching ratio between elastic and inelastic tunneling events (*25*). Because the main observation, i.e., the dramatic drop of light intensity when the tip is placed on top of the step edge, could be observed in every step edge regardless of its orientation, none of the previous studies considered the possibility that the light emitted from the picocavity could be anisotropic. Closer inspection of Figure 1B shows, however, that the intensity drop is not quite the same for step 1 and step 2, a fact that will become relevant later on.

As discussed above, raw electroluminescence intensities in STM are determined not only by the optical properties of the picocavity but also by electronic structure factors that consider the relative abundances of empty and occupied states in the tip and sample. Thus, the intensity drop in Figure 1 might arise from the different electronic structures of flat terraces versus step-edge positions. To unambiguously prove the origin of atomic-like contrast in STML, the electronic structure factor of the measured raw light emission spectra must be properly removed to unveil the pure optical properties of the picocavity, i.e., the radiative component of the Photonic Density of States (PhDOS) along the direction of optical detection. In short, as shown in our previous publication (22), to remove the electronic structure factor, one must divide the raw light emission spectra by the rate of inelastic tunneling, $R_{inel}(\hbar\omega, V_b)$, which can be experimentally evaluated from the $I_t(V_b)$ curves as $R_{inel}(\hbar\omega, V_b) \approx I_t(V_b - \hbar\omega/e)$, being $I_t$ the tunneling current, $V_b$ the bias voltage, $\omega$ the photon energy, $e$ the electron charge, and $\hbar$ the reduced Planck constant.

Figure 2A shows the measured tunneling current as a function of the bias voltage up to the set point voltage at which the STML spectrum has been recorded when the tip is placed on a flat terrace (red curve) or a step-edge (blue curve). Because the current used to stabilize the junction is the same in both cases, both curves cross at the set-point voltage. Importantly, however, the curves are quite different at every other voltage, revealing a significantly different electronic structure. Once the $I_t(V_b)$ data are converted to $R_{inel}(\hbar\omega, V_b)$ (Figure 2B), it can be observed that the change in curvature implies that the rate of inelastic tunneling is much greater when the tip is over the terrace than the step. This result indicates that the STML raw signal at the step will always be smaller than in the terrace just because fewer inelastic events can excite the LSP modes. The evolution of $R_{inel}(\hbar\omega, V_b)$ across the two monoatomic height steps over the black dotted line in Figure 1A is represented in Figure 2C. As for the raw STML, the color-coded image displays two abrupt drops, which appear at the positions of the steps, with a lateral extension of <1 nm each. Figure 2D shows the comparison of the spatial dependence of $R_{inel}(\hbar\omega, V_b)$ (blue curve) together with the raw light intensity at a photon energy of 1.81 eV (green curve) and the height profile of the STM tip (black curve). The topographic features of $R_{inel}(\hbar\omega, V_b)$ and the STML raw signal coincide with an accuracy only limited by the STM lateral resolution (~ picometer).

The experimental PhDOS obtained by dividing the raw light emission spectra by their corresponding rate of inelastic tunneling is shown in Figure 3. As shown in Figure 3B, a comparison with the raw STML spectra in Figure 1B demonstrates that most intensity changes are removed after considering the electronic structure factor. The huge decrease in intensity when the tip is placed atop a step edge has disappeared, and all spectra now have similar overall intensities. Thus, the atomically sharp contrast observed in STML of plasmonic systems mainly results from the underlying electronic structure factor, which is inherently convoluted to the raw luminescence. However, a closer inspection of the data shows that small changes in the radiative PhDOS at the steps remain, which can now be attributed to a purely optical effect. Figure 3A shows a color map of the measured PhDOS across the black dotted line in Figure 1A, traversing the two steps as a function of the photon energy and the tip position. Small changes in intensity are still noticeable: the intensity of the PhDOS of the left step is higher than in the flat terrace, and the right one is lower. When the STM tip is placed more than 1 nm away from the a step, the PhDOS is constant. Interestingly, in the vicinity of the left step, the PhDOS increases by 15%, decreasing by 10% when the tip is over the right step compared to the terrace. The observed changes in the radiative PhDOS are extremely local, having a lateral extension of <1 nm around each step. To emphasize the locality of the observed changes, Figure 3C shows the spatial dependence of the measured PhDOS at a fixed photon energy of 1.81 eV (green dotted cut of Fig 3A) plotted with the height profile of the scanned line. It should be noted that our results are independent of the sense in which the STM tip is scanned along the steps, ruling out that the observed effect could be an artifact related to a slow response of our feedback when crossing the step edge.

The fact that the measured PhDOS is appreciably modified by as much as 15% by a step of just one atom height (~2 Å) is a priori surprising. After ruling out that our observations result from local variations in the electronic density of states by properly removing its contribution from the raw STML spectra, which would modify the number of inelastically tunneling electrons and thus the detected raw light intensity, we can conclude that the effect must be purely optical. Moreover, the intensity differences between both steps must be related to changes in the orientation of the emitted radiation with respect to our collection angle. Notice that, by symmetry, the energy and intensity of the plasmonic modes of the picocavity formed between our tip and the sample when the tip is placed on top of the left step should be the same as when the tip is placed on the right step. However, the electric field distributions of the plasmonic modes are not necessarily identical, but they should be mirror images of each other through a plane perpendicular to the surface that contains the step edge. The radiative relaxation of the plasmonic modes, thus, could be different on both sides of such a plane. Because our collection angle is fixed (directed along the -x direction, see Figure 4A), our experiments give us information about the emitted radiation towards the half-space containing the upper terrace in the left step and towards the half-space containing the lower terrace in the right step, as schematically represented in Figure 3D.

To gain a deeper understanding of our experimental results, we conducted electromagnetic (EM) calculations of the STM picocavity, including the stepped silver surface. We employed the same modeling scheme presented in (*22*), based on the finite-element Maxwell's Equation solver implemented in Comsol Multiphysics. In agreement with experimental data, we modeled the STM tip as a Au sphere of radius 5 nm, placed at a vertical distance of 0.5 nm with respect to the upper terrace and a perfectly vertical step with a height of 0.2 nm. Radiated PhDOS spectra for different tip-step lateral positions were obtained by integrating the time-averaged Poynting vector in the far-field, within solid angles mimicking the experimental collection lens. As expected, our EM simulations indicate that the radiated power is isotropic when the STM tip is over the flat terrace. On the contrary, when the tip is atop a monoatomic step of just 2 Å height, the radiated power becomes strongly anisotropic, displaying a cardioid shape with a very sharp minimum in the direction perpendicular to the step edge towards the half-space containing the lower terrace. The simulated angular distribution of the emitted light as a function of the azimuthal angle at the photon energy that yields maximum radiated power is shown in the inset of Figure 4A. The blue circumference corresponds to the simulated angular distribution when the tip is over the terrace, i.e., 3 nm away from the step, and the black cardioid when the tip is atop the step, i.e., 0.25 nm from its origin (see Figure S1 for the criterion for choosing the origin of the step). Depending on the azimuthal angle, the radiated power is higher/lower when the tip is over the terrace/step, yielding a higher/lower radiative PhDOS along that direction, strongly reminiscent of our experimental results.

Knowing that our collection direction is -x, we can calculate the azimuthal angle that corresponds to the emission that we experimentally measure. Figure 4A shows an STM image of a clean Ag(111) surface with steps at different azimuthal angles referenced to the direction along our collection lens. Note that the orientations at which the steps appear are not random and are determined by the high symmetry directions of the crystal. Here, straight step edges were created by gently crashing the STM tip into the substrate, after which the surface was reconstructed. We have measured the PhDOS across the four steps in Figure 4A (see supplementary figure S3), displaying three different orientations. Steps 1 and 4 (marked in red and green, respectively) are equivalent, forming an azimuthal angle of $22^0$ referenced to our collection direction. Steps 1 and 2 (marked in red and pink, respectively) are mirror symmetric, forming angles of $22^0$ and $202^0$, respectively. Step 3 (marked in blue) forms an angle of $142^0$. The experimental values of the integrated PhDOS for the four steps are represented by solid-colored dots in Figure 4B. The gray shaded area represents the lower terrace of a step, and the orange shaded area represents the upper one. Following the theoretically predicted anisotropic shape of the PhDOS at the steps, the solid cyan line corresponds to the best possible fit to a cardioid-like function $r = r_1 + (r_2 - r_1) \times \left|\sin\left((\pi - \theta)/2\right)\right|$, where $r$ and $\theta$ are the polar coordinates (radius and azimuthal angle), and $r_1 = 0.8$ and $r_2 = 1.2$ are the fitting parameters (radii of the two circumferences forming the cardioid). The green dotted circumference, representing isotropic emission, has been included as a guide to the eye to emphasize the directional nature of the emitted light at the steps. Notice the agreement between experiment and theory: when the lower terrace of a step is oriented to our collection (gray shaded area), the data points lie inside the green circumference, whereas when it is the upper terrace (orange shaded area), the data points lie outside. This means that the PhDOS is higher/lower on the terrace than in the step when the step is in an upward/downward orientation with respect to our collection direction. However, a quantitative comparison between experiment and theory is beyond the scope of our simulations due to the two main simplifications of our model. First, it is assumed that the step is perfectly vertical, not considering the spill out of electrons due to the Smoluchowski effect. Second, the vertical displacement of the STM tip while traversing the step is not taken into account, the distance is fixed to 0.5 nm. In addition to the directionality of the emitted light, there is a spectral reshaping of the PhDOS at the steps, displaying a sigmoidal-shaped curve with its midpoint at the maximum of the plasmonic emission, in excellent agreement with our simulations (see supplementary Figure S2).

**Discussion**

The directionality of nanoscale light is a key element for multiple applications relying on quantum light sources. Arrangements of nanostructures can control the directionality of light by far-field interference of different multipolar modes of its constituents when excited by laser irradiation. In nanoantenna-based devices, this can be achieved by engineering the size and shape of the nanoantennas. Alternatively, plasmonic light can be beamed by focusing high-energy electrons (~ keV) in cathodoluminescence experiments or low-energy electrons (~ eV) in STM on an individual nanostructure, where its geometry and excitation position in the few-nanometer scale also determines the emission direction (*29, 30*). The mentioned strategies aiming to control the directionality of nanoplasmonic sources rely on imprinting far-field changes of the emitted light by modifying the near field at relatively large scales (few-nanometers), compared to the investigations carried out in this work.

Our results demonstrate that monoatomic steps with a height of only 2 Å significantly influence the near-field of plasmonic nanoantennas, thus producing non-negligible effects that can be detected in the far-field. The measured changes in the PhDOS of our tip-sample picocavities are confined to a lateral extension of ~1 nm around each step. The physical origin of the extreme spatial sensitivity found in our experiments is the ultraconfinement of electromagnetic fields inside such picocavities and the exponential dependence of the tunneling current on the tip-sample distance, which may be regarded as an extremely local excitation source of LSP modes. EM calculations of the STM nanocavity, including a monoatomic height step, show that the experimentally detected changes in the PhDOS result from tilting the emission angle at the steps. Although the picocavity LSP modes do not have a purely vertical dipolar character, they acquire an in-plane dipole moment due to the presence of the step. The tilting angle of the emitted light is such that a 15% increase/reduction in the intensity of the radiative PhDOS is observed in the far field. It is important to recall that to access the pure optical properties of the nanocavity, i.e., the radiative PhDOS, one must remove the electronic structure factor from the raw light emission spectra by dividing them by the rate of inelastic tunneling. Otherwise, the obtained spectra are predominantly dominated by the electronic structure factor, hindering the true optical properties of the system.

The outcomes of our investigations suggest that atomic-like defects might play a relevant role in determining the nm-averaged response of nanoscale light sources. This may open new avenues for designing novel picophotonic devices with patterned structures at the atomic scale for controlling light emission properties, with implications in multiple fields, such as sensing, quantum information, and energy storage, to mention a few.

**Materials and Methods**

**Experimental setup**
The experiments were conducted in a custom-designed low-temperature STM from Omicron operating at liquid helium temperature (4 K) and in ultrahigh vacuum conditions ($P < 10^{-10}$ mbar). The STM tips were made of gold. A 0.40 mm diameter wire was electrochemically etched to sharpen the tips in a 50/50 solution of ethanol and HCl (37% purity). Clean Ag(111) surfaces were prepared by alternating cycles of $Ag^+$ sputtering at an energy of 1.5 keV followed by thermal annealing at 500 $^0$C. The straight step edges following the high symmetry directions of the substrate were created by gently crashing the STM tip onto the substrate. The electroluminescence signal was first collected by an *in-situ* lens of focal length $f = 15\ mm$ placed 15 mm away from the tip-sample junction and then guided in free space to an optical spectrometer (Andor

Shamrock 500), equipped with a Peltier-cooled charge-coupled device (CCD). All the presented spectra are background corrected but are not corrected by the efficiency of the setup.

**Data acquisition**
To confront the electroluminescence data with the rate of inelastic tunneling, we recorded STML data and STS curves at each tip position. The feedback loop was kept closed for the STML data to ensure a constant tunneling current during acquisition. For the STS data, the feedback was opened at the same initial setpoint conditions as for STML, and then the bias voltage was swept from the stabilization bias at which the STML data was acquired down to zero. After acquiring an STML spectrum, an STS spectrum at the same position was acquired. To record the next STML spectrum, the feedback was turned on to keep a constant tunneling current while the tip moved. We created a Python script using MATE-for-Dummies (*31*) to govern the routine for data acquisition.

**Electromagnetic simulations**
The numerical calculations were carried out using the frequency-domain finite element solver of Maxwell's Equations implemented in the commercial software Comsol Multiphysics. A full 3D simulation domain was required to accommodate the Au 5 nm radius nanosphere modeling the STM tip (*27*) and the Ag flat substrate, decorated by a 50 nm radius, 0.2 nm height half disc modeling the atomic-sized step. The height and radius of the total simulation volume was ~3λ (excluding perfect matching layers). A conformal mesh distribution accounted for the electromagnetic field propagation from the picometric antenna gap, driven by a point-like dipole at its center to the circular-shaped area where the time-averaged Poynting vector was integrated. This was located ~1λ away from the dipolar source, and its dimensions were set to mimic the solid angle covered by the collection lens in the experiments. The convergence of numerical results against the mesh size and distribution was checked. All permittivities were taken from (*32*).

**Acknowledgments**
A.M-J acknowledges funding from HORIZON-MSCA-2022-PF-01-01 under the Marie Skłodowska-Curie grant agreement No. 101108851. A.I.F.-D. acknowledges funding from the Spanish Ministry of Science,

Innovation and Universities through Grants PID2021-126964OB-I00 and TED2021-130552B-C21, as well as the European Union's Horizon Programme through grant 101070700 (MIRAQLS). R.M. and R.O. acknowledge financial support from the Spanish Ministry for Science and Innovation (Grants PID2020-113142RB-C21, PLEC2021-007906, and PID2021-128011NB-I00). Both IMDEA Nanoscience and IFIMAC acknowledge support from the Severo Ochoa and Maria de Maeztu Programmes for Centres and Units of Excellence in R&D (MICINN, Grants CEX2020-001039-S and CEX2018-000805-M). R.O. acknowledges support from the excellence program for University Professors, funded by the regional government of Madrid (V PRICIT). The authors acknowledge the support from the "(MAD2D-CM)-UAM" project funded by Comunidad de Madrid, by the Recovery, Transformation and Resilience Plan, and by NextGenerationEU from the European Union. D.M acknowledges funding from "Ayuda PRE2022-101740 financiada por MCIN/AEI/10.13039/501100011033 y por el FSE+". M.V acknowledges funding from "Ayuda PRE2022-104827 financiada por MCIN/AEI/10.13039/501100011033 y por el FSE+".

**Author contributions:**
D.M, M.V, and O.J contributed equally to this work. D.M, O.J, M.V, K.L, and A.M-J performed the experiments. D.M, M.V, A.M-J, and R.O analyzed the data. A.I.F.D performed theoretical modelling and simulations. R.O conceived the project. A.M-J wrote the first version of the manuscript. All authors discussed the data and contributed to the final version of the manuscript.

**Competing interests:**
The authors declare no competing interests.

**Data and materials availability:**
All data and code used in the analyses are available from the corresponding author upon reasonable request.

**Figures and Tables**

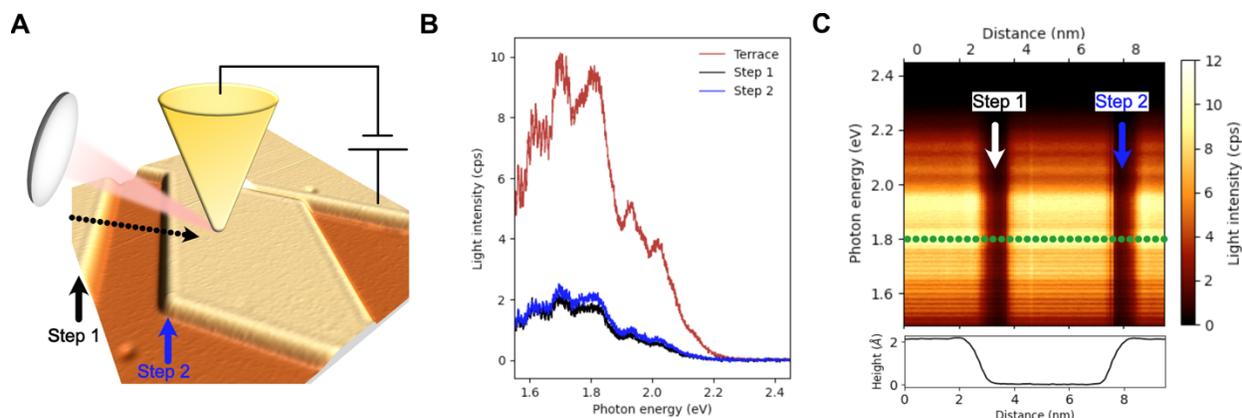

**Figure 1. Raw light emission across a monoatomic height step in an STM picocavity. (A)** Schematic representation of the experimental setup. A lens near the STM tunnel junction collects the plasmonic luminescence of the tip-sample LSP modes. The black dotted line denotes the trajectory of the tip across two parallel steps (steps 1 and 2, respectively) over the Ag(111) surface. STM image acquisition settings: size = 25 nm x 25 nm, $I_{sp}$ = 10 pA, $V_{sp}$ = 1.0 V. **(B)** Raw light emission spectra when the STM tip is on top of a monoatomic height step, blue and black curves, step 1 and step 2 in (A) respectively, and over the flat terrace, red curve. Acquisition settings: integration time = 120 seconds, $I_{sp}$ = 350 pA, $V_{sp}$ = 2.6 V. **(C)** Spatially resolved raw light intensity (color-coded) as a function of the photon energy and the tip displacement over the black dotted line in (A). Acquisition settings: integration time/spectrum = 120 seconds, $I_{sp}$ = 350 pA, $V_{sp}$ = 2.6 V; number of spectra = 120. Bottom panel: height profile corresponding to the trajectory followed by the STM tip in the constant current operation mode demonstrating that the steps are one atom in height.

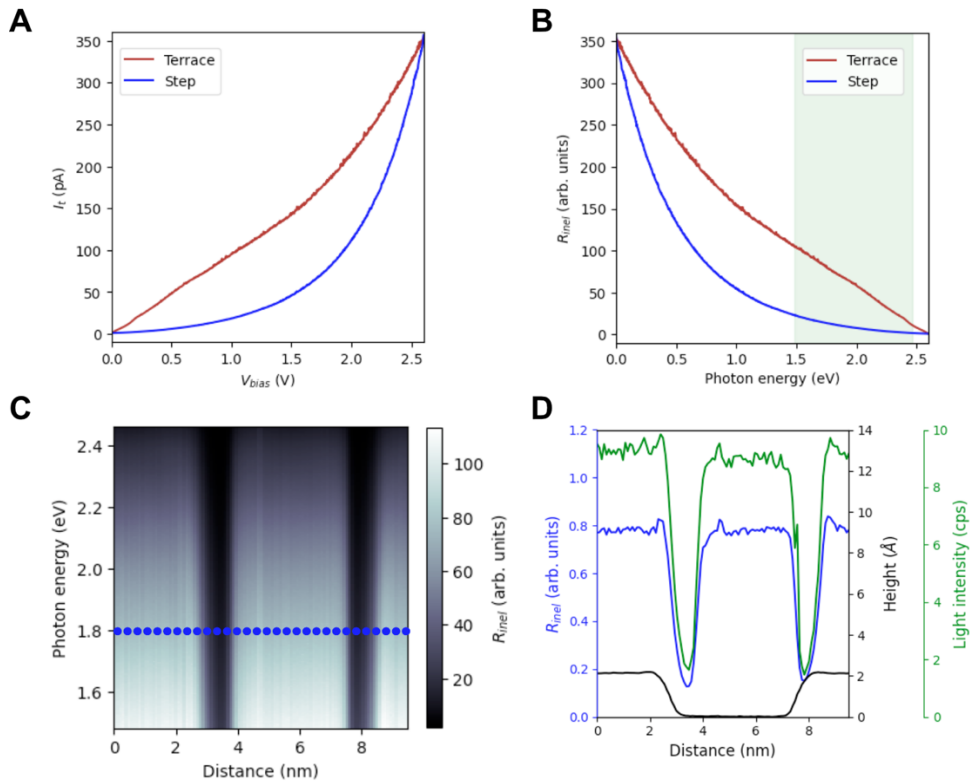

**Figure 2. Rate of inelastic tunneling across a monoatomic height step. (A)** Tunneling current as a function of the bias voltage, $I_t(V_b)$ curve. Red curve: STM tip over the clean terrace. Blue curve: STM tip atop a monoatomic height step. The tunneling current feedback loop was open during the acquisition. The initial setpoint values determining the tip-sample distance were the same as for the STML curves: $I_{sp}$ = 350 pA, $V_{sp}$ = 2.6 V. **(B)** Rate of inelastic tunneling ($R_{inel}(\hbar\omega, V_b)$) obtained from the curves in (A). The green shaded area represents the photon energy window over which the STML data have been acquired. **(C)** Color-coded representation of $R_{inel}(\hbar\omega, V_b)$ as a function of the tip displacement over the black dotted line in Figure 1A. The image corresponds to a vertical stacking of 120 curves. The tunneling current feedback loop was enabled between consecutive curves while the tip moved to maintain a constant current setpoint over the scan line. **(D)** Black curve: height profile of the scanned line in the constant current mode of the STM. Blue curve: $R_{inel}(\hbar\omega, V_b)$ across the scanned line at a photon energy of 1.81 eV. Green curve: raw light intensity at a photon energy of 1.81 eV across the scanned line.

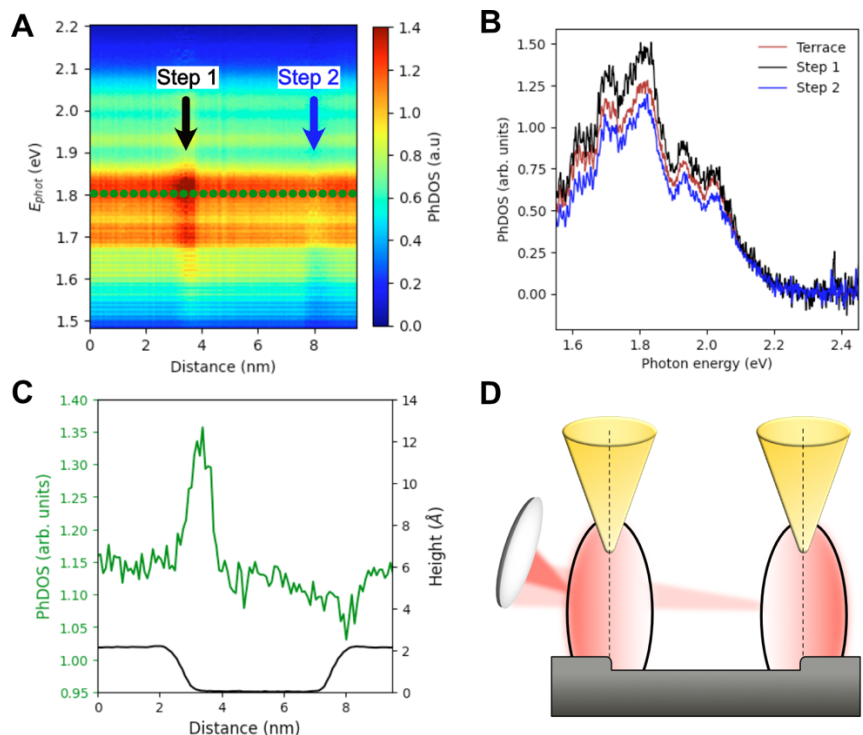

**Figure 3. The photonic density of optical states across a monoatomic height step. (A)** Experimentally obtained radiative PhDOS along the black dotted line in Fig 1A. After normalization of the STML raw spectra by $R_{inel}$, the abrupt changes in light intensity in the proximity of the monoatomic height steps are removed. Still, some small differences remain, originating from local modifications of the true optical properties of the tip-sample picocavity. An enhancement/decrease of the PhDOS at the step positions compared to the terrace is observed. **(B)** Spectral distribution of the PhDOS when the STM tip is over the terrace (red curve), over step 1 (black curve), and over step 2 (blue curve). Each curve corresponds to a vertical cut of the color plot in (A). The PhDOS of the picocavity formed by the STM tip and the downward step (step 1) is noticeably higher than for the upward step (step 2). **(C)** Green curve: PhDOS across the green dotted line in (A) at a photon energy of 1.81 eV. Notice that the PhDOS locally increases/decreases in the proximity of the two steps referenced to the flat terrace, confined to a lateral extension <1 nm. Black curve: height profile of the scanned line in the constant current mode of the STM. **(D)** Schematic illustration of directional emission for two mirror-symmetric steps. The ellipses represent the distribution of the plasmonic modes of the tip-sample picocavity near the step edges. Only the left part of each ellipse appears red-shaded, indicating the half-plane of collection defined by our lens, placed along the -x direction (see Figure 4A for definition of the collection direction). Notice that for the left step, the half-space of the emitted light contains an upper terrace, whereas for the right step, it contains a lower terrace.

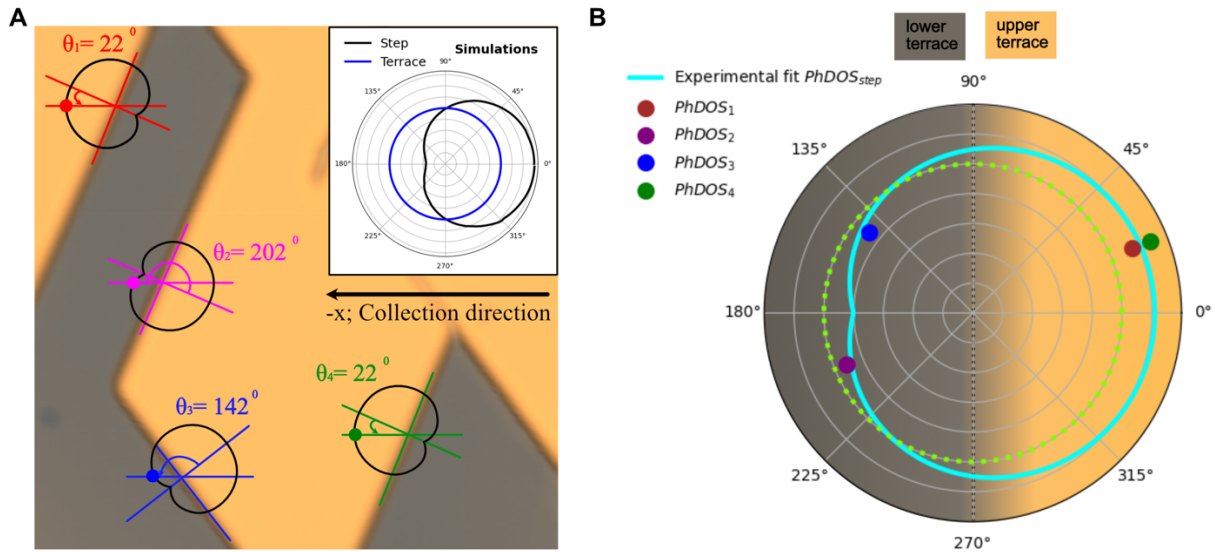

**Figure 4. Directionality of plasmonic luminescence for picocavities with a monoatomic height step.** (**A**) STM image of a Ag(111) surface with steps at different orientations. STM image acquisition settings: size = 25 nm x 25 nm, Isp = 10 pA, Vsp = 1.0 V. The angles formed between the line perpendicular to each step and our collection direction (-x) are annotated. Inset: EM simulations of the PhDOS when the tip (sphere) is placed atop a monoatomic height step (black curve), and over the flat terrace (blue curve), as a function of the azimuthal angle in polar coordinates. (**B**) Measured angular distribution of the emitted light. The colored dots correspond to the integrated PhDOS when the STM tip is located at each step, referred to each step's frame of reference, where the upper terrace is at the right (orange shaded background), and the lower terrace is at the left (gray shaded background). Red, pink, blue, and green dots correspond to step 1, 2, 3, 4, respectively. The experimental data at the three azimuthal angles defined by the orientations of the steps do not follow a circumference, green dotted circle representing isotropic emission at the terrace, implying directional emission. The cyan line is a fitting of our experimental data to a cardioid-like function.

**Supplementary Materials**

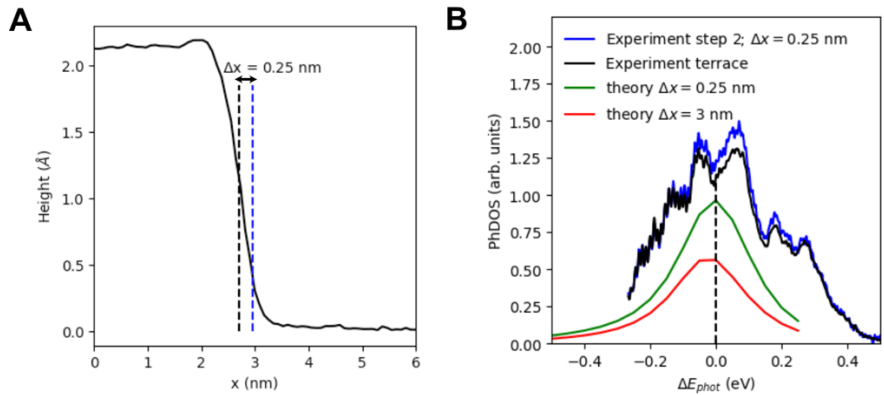

**Figure S1. Referencing the PhDOS to the photon energy of the centroid of the plasmonic emission.** (**A**) Height profile over a monoatomic height step acquired in the constant current mode of the STM. The dashed black vertical line at the center of the slope represents the origin chosen for referencing the lateral distance between the STM tip and the step. The dashed blue vertical line at $\Delta x=0.25$ nm indicates the lateral position of the STM tip chosen for comparing our EM simulations and experimental data. The selected value of $\Delta x=0.25$ nm is arbitrary, and defines a distance considered as the tip being over the step. When $\Delta x$ is beyond 3nm the effect of the step is negligible and the PhDOS is the same as when the tip is over a perfectly flat terrace. (**B**) The blue and black curves correspond to experimental data with the STM tip over step 2 and over the terrace,

respectively, in Figure 1A. The black dashed vertical line indicates the centroid of the PhDOS at $E_{phot} = 1.75$ eV. The green and red curves correspond to the simulated PhDOS for tip positions at Δx=0.25 nm (step), and Δx=3 nm (terrace), respectively.

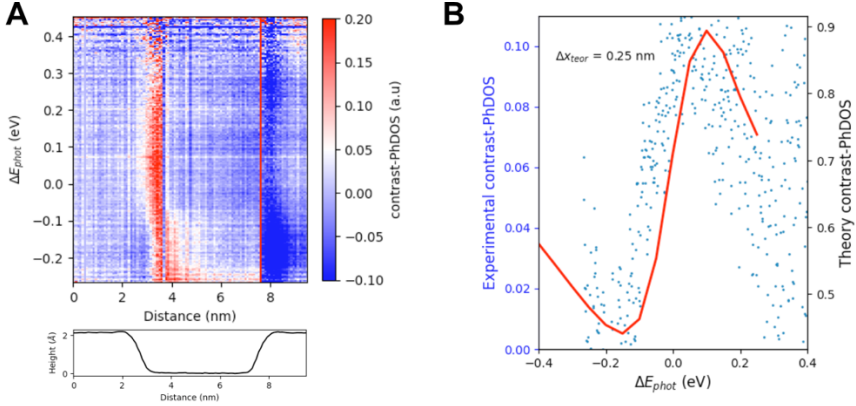

**Figure S2. Spectral reshaping of the radiative PhDOS near a monoatomic height step edge.** To discriminate if the modifications of the PhDOS are just a change in the global intensity or whether the spectral distribution has also changed, we define a spectrally-resolved contrast factor as $(PhDOS_{step}(\hbar\omega) - PhDOS_{terrace}(\hbar\omega))/PhDOS_{terrace}(\hbar\omega)$. **(A)** Spectrally-resolved contrast factor of the PhDOS, from now on contrast-PhDOS, as a function of the relative photon energy and tip position along the black dotted line in Figure 1A. The contrast plot highlights the spectral reshaping of the radiative PhDOS when the tip is over each step. The higher/lower PhDOS at the two steps is now visualized as a positive/negative quantity in the contrast map. Notice that each curve is referenced to the photon energy corresponding to the maximum intensity ($\Delta E_{phot}$). **(B)** Spectral distribution of the contrast-PhDOS when the STM tip over a step edge (Δx=0.25 nm). Blue dots: experimental data. Red curve: EM simulations. Not all wavelength components are enhanced in the same way. The contrast-PhDOS has a sigmoidal shape with its center at the photon energy of the maximum of the plasmonic emission, $\Delta E_{phot} = 0$.

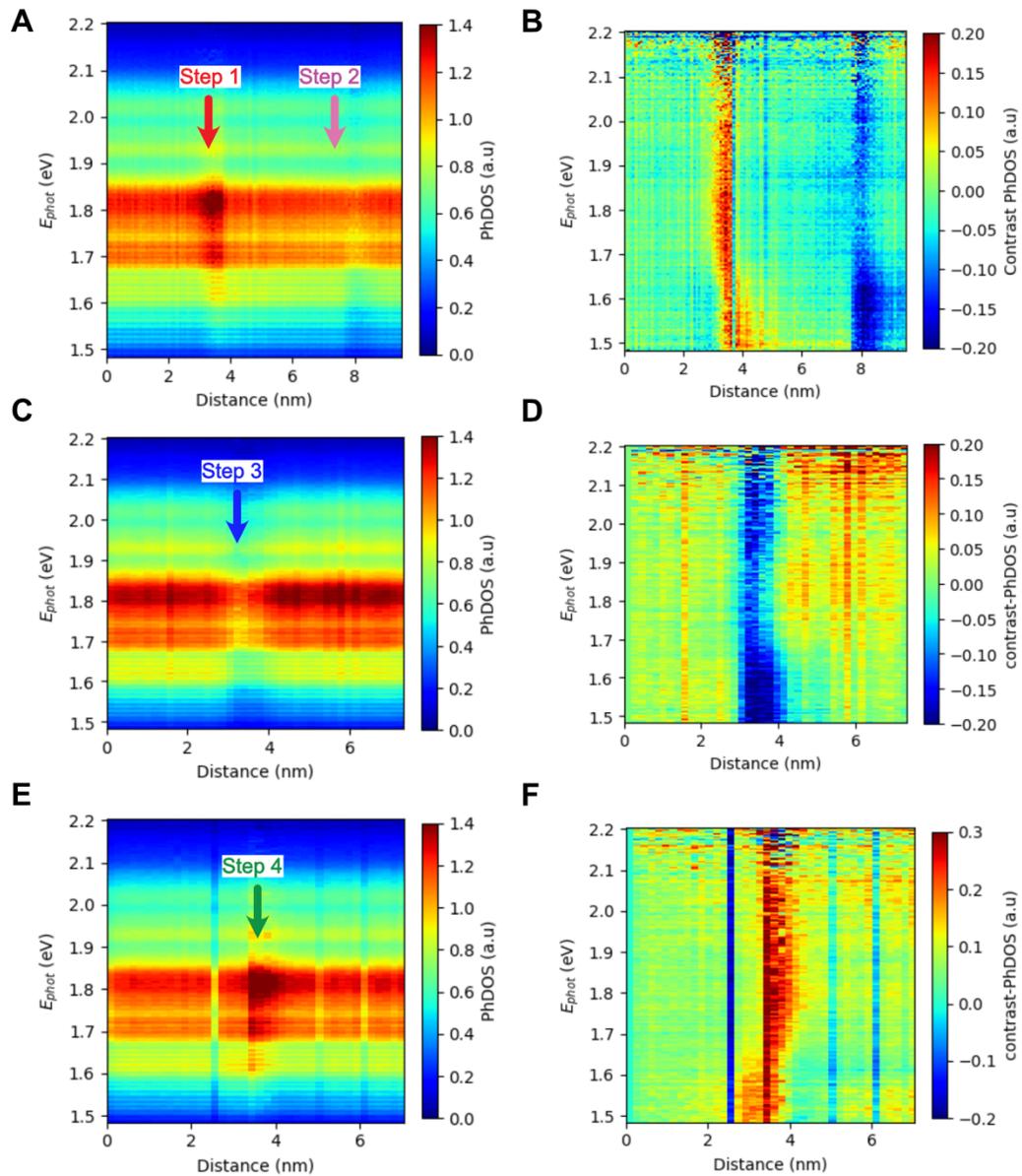

**Figure S3. PhDOS for different orientation steps. (A)** PhDOS as a function of the photon energy and tip position along steps 1 and 2 in Figure 4A. **(B)** Corresponding contrast-PhDOS of (A). **(C)** PhDOS as a function of the photon energy and tip position along step 3 in Figure 4A. **(D)** Corresponding contrast-PhDOS of (C). **(E)** PhDOS as a function of the photon energy and tip position along step 4 in Figure 4A. **(F)** Corresponding contrast-PhDOS of (E).